\documentclass[graybox]{svmult}

\usepackage{mathptmx}       
\usepackage{helvet}         
\usepackage{courier}        
\usepackage{makeidx}         
\usepackage{graphicx}        
\usepackage{multicol}        
\usepackage{amsmath,amssymb,amsfonts}        
\makeindex             

\usepackage{mathrsfs}
\usepackage[utf8]{inputenc}
\usepackage[russian,english]{babel}
\usepackage{slashed}

\usepackage{color}
\definecolor{dark-gray}{gray}{0.20}
\definecolor{gray}{gray}{0.30}
\definecolor{light-gray}{gray}{0.80}
\definecolor{dark-red}{rgb}{0.7,0,0}
\definecolor{dark-green}{rgb}{0.1,0.4,0}
\definecolor{dark-blue}{rgb}{0.3,0.3,0.7}
\definecolor{light-blue}{rgb}{0.8,0.8,1}
\definecolor{blue}{rgb}{0,0,1}
\definecolor{red}{rgb}{1,0,0}
\definecolor{green}{rgb}{0,1,0}

\usepackage{hyperref}
\hypersetup{
	colorlinks=true,
	linkcolor=dark-blue,
	citecolor=dark-red,
	urlcolor=dark-blue,
	linktoc=page
}

\def\cL{{\cal L}}

\def\cN{{\cal N}}

\def\cR{{\cal R}}

\def\cV{{\cal V}}

\def\cX{{\cal X}}

\def\i{{\rm i}}

\newcommand{\be}{\begin{equation}}
\newcommand{\ee}{\end{equation}}
\newcommand{\bea}{\begin{eqnarray}}
\newcommand{\eea}{\end{eqnarray}}

\begin{document}

\title*{Equivariant localization and gluing rules\\ in 4d $\cN=2$ higher derivative supergravity}

\author{Kiril Hristov}

\institute{Faculty of Physics, Sofia University, J. Bourchier Blvd. 5, 1164 Sofia, Bulgaria \\
INRNE, Bulgarian Academy of Sciences, Tsarigradsko Chaussee 72, 1784 Sofia, Bulgaria}

\maketitle

\abstract{
We complete the proof of the conjecture in \cite{Hristov:2021qsw} regarding the form of gravitational building blocks in higher derivative supergravity and examine the related gluing rules. The derivation, based on certain assumptions, combines the BVAB equivariant localization formula (on non-compact spaces) and the direct evaluation of the action on the Omega background in AdS$_4$, \cite{Hristov:2022plc}. Utilizing the superconformal gravity formalism and avoiding equations of motion ensures our results are valid at any perturbative order in the Newton constant. As an application, we illustrate the holographic predictions for supersymmetric M2-brane partition functions. We conclude with a to-do list of related calculations in effective supergravity and string theory, aiming for full quantum control of BPS backgrounds.}
%
%


\section{Introduction}
\label{sec:intro}
The application of supersymmetric localization à la Pestun \cite{Pestun:2007rz} to supergravity backgrounds was initiated in \cite{Dabholkar:2010uh} for black holes in ungauged supergravity, which allows for arbitrarily small internal manifolds in string theory. This was extended to gauged supergravity in \cite{Hristov:2018lod}, where Kaluza-Klein modes cannot be disregarded due to the fixed volume of the compactification manifold. Both ungauged and gauged cases required additional assumptions to constrain the metric and ensure supergravity's perturbative quantum viability. A more mature approach, inspired by localization à la Nekrasov \cite{Nekrasov:2002qd}, began with \cite{BenettiGenolini:2019jdz,Hosseini:2019iad} and was rigorously developed in \cite{BenettiGenolini:2023kxp,Martelli:2023oqk}.~\footnote{Even though the latter reference is closely related to the main results of this work, the approach is geometric and the starting point is string theory rather than effective supergravity. On the other hand it is potentially more fundamental, as commented in the last section.} It employs the equivariant localization techniques of Duistermaat-Heckman \cite{Duistermaat:1982vw} and Berline-Vergne-Atiyah-Bott (BVAB) \cite{berline1982classes,Atiyah:1984px} based on a special $U(1)$ action generated by susy, and eschews the need of explicit metrics.

The approaches of Pestun and Nekrasov are connected via the observation that for backgrounds with fixed points under the $U(1)$ action, the result for the partition function simplifies to a sum over isolated fixed point contributions. It is then useful to consider the simplest background with a fixed point, known as the Omega background \cite{Nekrasov:2002qd, Nekrasov:2003rj}. The Nekrasov partition function, derived from this background, serves as a building block. Together with a set of \emph{gluing rules} that describe the relation of geometric (equivariant) and physical (Coulomb branch) parameters at each fixed point, one can compute all other partition functions \cite{Nekrasov:2003vi}. This idea was revisited in supergravity for the formulatation of the conjecture in \cite{Hristov:2021qsw} for the gravitational building blocks of general higher derivative supergravity, whose proof we present here.

The main idea in \cite{Hristov:2021qsw} was to generalize the gravitational building blocks of \cite{Hosseini:2019iad} to off-shell supergravity with higher derivative invariants. The conjectured form of a single block as an infinite expansion governed by the equivariant parameters was inspired by examples such as \cite{LopesCardoso:1998tkj,Hristov:2016vbm,Bobev:2020egg} and more rigorously supported by the calculation in \cite{Hristov:2022plc} for the gauged supergravity version of the Omega background, called Omega-AdS$_4$.~\footnote{It is closely related to the supergravity dual of the squashed three-sphere of \cite{Martelli:2011fu}.} Utilizing equivariant localization for the supergravity action, as detailed in \cite{BenettiGenolini:2023kxp, BenettiGenolini:2023ndb}, we are now able to formally complete the proof of the conjecture in \cite{Hristov:2021qsw}. To assemble the proof, we utilize the BVAB fixed point formula together with the major shortcut of the direct Omega background evaluation in \cite{Hristov:2022plc} that utilizes holographic renormalization, \cite{Skenderis:2002wp}.

The logic of our proof is novel to the supergravity literature and is applicable in a broader range of circumstances, listed in the last section. Here, it enables us to generalize the results of \cite{BenettiGenolini:2023kxp} by incorporating arbitrary higher derivative corrections to the action and including arbitrary matter couplings, all while keeping calculations entirely off-shell. The latter feature is of fundamental importance from supersymmetric localization point of view. However, as we also comment later, it would be very interesting to repeat the proof following the steps of \cite{BenettiGenolini:2023kxp, BenettiGenolini:2023ndb} without any shortcuts. This requires the generalization of the holographic renormalization procedure in the off-shell formalism.

Another very important point that we elaborate upon, already noticed in \cite{Hristov:2021qsw}, is that the \emph{gluing rules} for a given background depend only on the topology of the background and the field content of the theory and are independent of the Lagrangian, such that higher derivative corrections alter only the gravitational blocks but not the gluing rules. Off-shell supergravity allows us an even stronger handle on the gluing: we can distinguish and fully independently determine the rules for the equivariant parameters, determined entirely within the gravity multiplet, and the Coulomb branch parameters stemming from the addition of any extra matter multiplets.

The rest of this note is organized as follows. In the next section, we discuss equivariant localization and the core idea of our proof using the Omega background in general terms, applicable to any even-dimensional theory. In Section \ref{sec:sphere}, we examine the key aspects of the off-shell supergravity formalism and the conditions imposed by supersymmetry. Section \ref{sec:omega} presents the calculation from \cite{Hristov:2022plc} and completes the proof of the conjecture in \cite{Hristov:2021qsw}. Sections \ref{sec:setup} and \ref{sec:m2} cover the gluing rules for black hole geometries and apply the formalism to some interesting M2-brane partition functions, respectively, connecting our discussion to the results of \cite{Hristov:2022lcw}. We finish with a to-do list of related topics, outlining the author's short-term research interests.

\section{Equivariant localization in supergravity}
\label{sec:equiloc}
Here we mostly follow the logic of  \cite{BenettiGenolini:2023kxp, BenettiGenolini:2023ndb}, stating our main assumptions along the way. Let us consider a generic Euclidean supergravity action for a metric $g_{\mu\nu}$, a number of gauge fields $W^I_\mu$ and scalars $X^\alpha$ of the form
\be
	I (M) =  \int_M \cL(g_{\mu\nu}, W^I_\mu, X^\alpha, ...)+ \int_{\partial M} \cL_\text{bdry} (h_{i j}, W^I_i, X^\alpha, ...) \ ,
\ee
where $M$ is an asymptotically locally (Euclidean) AdS, the metric $h_{ij}$ is the induced metric on the boundary $\partial M$, and the dots denote unspecified fermionic fields. According to the standard idea of holographic renormalization for asymptotically locally (Eculidean) AdS geometries, \cite{Skenderis:2002wp}, the divergent boundary contributions of the first integral precisely cancels with the second integral. Furthermore, we assume that the counterterms are susy invariant in such a way as to make sure the following equivariant localization theorem holds.~\footnote{In other words, we assume the principles of \emph{exact} holography hold for the considered action. It is not a mathematical certainty that equivariant localization must hold on non-compact spaces, but this is a natural expectation from AdS/CFT perspective.} 

Let us further assume we are in even dimensions, $d = 2 n$, such that after the procedure of holographic renormalization we can write down the action as
\be
	I (M) = \int_M \Phi_{2 n} (g_{\mu\nu}, W^I_\mu, X^\alpha, ...)\ ,
\ee
where $\Phi_{2 n}$ is automatically a top form on $M$. We are going to be interested in supersymmetric backgrounds, and generically assume at least two preserved supercharges. The ability of the background to admit a pair of conformal Killing spinors  (see below for the explicit details in 4d $\cN=2$ superconformal gravity), would then imply the existence of a specific $U(1)$ action on it. Even if not strictly needed, we are going to see that the special $U(1)$ is a conformal isometry of $M$. We denote it by the vector $\xi$,~\footnote{Conformally equivalent backgrounds in the off-shell formalism automatically have the same supersymmetric properties. From the arguments presented here it also follows that their action is the same, such that we can always choose a Weyl representative of a given class where $\xi$ is a precise isometry.} which allows us to define the equivariant exterior derivative
\be
	{\rm d}_\xi := {\rm d} - \iota_\xi\ ,
\ee
such that it squares to (minus) the Lie derivative, ${\rm d}_\xi^2 = - \cL_\xi $. We can then construct the equivariant completion of the form $\Phi_{2 n}$,
\be
\label{eq:multiform}
	\Phi := \Phi_{2n} + \Phi_{2 (n-1)} + ... + \Phi_2 + \Phi_0\ ,
\ee
by solving the recursive differential equations
\be
\label{eq:deq}
	\iota_\xi \Phi_{2 k}= {\rm d} \Phi_{2 (k-1)}\ , \quad \forall k \in \{ 1, 2, ... n \}\ ,
\ee
where the subscript denotes the degree of the form. In other words, finding all $\Phi_{0, 2, .., 2 (n-1)}$ from $\Phi_{2 n}$ gives us an equivariantly closed polyform, 
\be
	{\rm d}_\xi \Phi (g_{\mu\nu}, W^I_\mu, X^\alpha, ...)  = 0\ ,
\ee
whose top-form coincides with the original expression $\Phi_{2 n}$. Since lower forms do not contribute directly to the integral over $M$, we can simply rewrite the action as
\be
\label{eq:preequiloc}
	I (M) =  \int_M \Phi_{2 n}= \int_M \Phi \ .
\ee
Note that we are not assuming anything else about the background, except that it preserves supersymmetry and thus admits the vector $\xi$. On the other hand, at this stage we need to assume the existence of a solution to \eqref{eq:deq}, as shown explicitly in \cite{BenettiGenolini:2023kxp,BenettiGenolini:2023ndb} for a number of supergravity actions based on supersymmetric identities. As shown there, the expression for $\Phi$ is in general non-unique as expected for solutions of differential equations, but all choices lead to the same integral $I(M)$, which is what ultimately interests us. 

{\bf Fixed points.} An important assumption is that we are looking only at backgrounds exhibiting isolated fixed points on $M$ under the action of $\xi$, labeled by $\sigma$.~\footnote{Notice that so far we followed the logic and main assumptions of \cite{BenettiGenolini:2023kxp,BenettiGenolini:2023ndb}. However, our aim here is more restrictive and we only look at fixed points of $\xi$, disallowing fixed submanifolds of higher dimension that equivariant localization allows for in general.} Using the BVAB localization formula, \cite{berline1982classes,Atiyah:1984px} (see e.g.\ \cite{Pestun:2016qko} for a physics oriented review), specified on fixed points of $\xi$, we can rewrite the integral simply as a sum over fixed points, 
\be
	I (M) = \sum_{\sigma}\, \frac{\Phi_0 |_{\sigma}}{e (TM) |_{\sigma}}\ ,
\ee
where $e (TM)$ is the equivariant Euler class of the tangent plane at the fixed point. 

In $d = 2 n$, the tangent space at any of the fixed points is $\mathbb{C}^n$, which we can parametrize in polar coordinates with $n$ radial coordinates $r_i$ and as many angular variables $\varphi_i$,
\be
	{\rm d} s^2_{\mathbb{C}^n} = \sum_{i=1}^n\,  ({\rm d} r_i^2 + r_i^2 {\rm d} \varphi_i^2)\ ,
\ee
such that locally the $U(1)$ vector $\xi$ takes the form of an arbitrary linear combination of the available $U(1)^n$ rotations,
\be
	\xi = b_1\, \partial_{\varphi_1} + ... + b_n\, \partial_{\varphi_n} = \sum_{i=1}^n  b_i\, \partial_{\varphi_i}\ ,
\ee
where $b_i$ are called equivariant parameters. Using the fact that the equivariant Euler class is the Pfaffian of the curvature two-form, see e.g.\ \cite{Bobev:2015kza}[App. A], we find
\be
	e (TM) = \prod_{i=1}^n\, b_i\ .
\ee

Depending on the topology of the background $M$, the number of fixed points $\sigma$ and the relation between their corresponding equivariant parameters will be different. We thus arrive at the simplified formula for the supergravity action,
\be
\label{eq:equiloc}
	I (M) = \sum_{\sigma}\, \frac{\Phi_0 |_{\sigma}}{\prod_{i=1}^n\, b_i |_{\sigma}}\ .
\ee

Note that the resulting bottom-form $\Phi_0$ is in general a scalar function of all background fields. Therefore we expect $I (M)$ to depend on a number of scalar parameters, no matter whether they are the physical scalars or some composite objects. From the analogy with Nekrasov's partition function, which can be thought of as the full non-perturbative generalization of $\Phi_0$ for 4d $\cN=2$ supersymmetric field theories, we are going to call them the Coulomb branch parameters, and distinguish them from the equivariant parameters. However, the distinction is (for now) purely linguistic since $\Phi_0$ itself is no longer the original action that allows a clear interpretation of different terms.

We thus arrive at the main physics corrolary, following from the mathematical derivation (under the stated set of assumptions) of \eqref{eq:equiloc}. For a background with fixed points under $\xi$, the only information we need is the explicit form of $\Phi_0$, ignoring \emph{all} other higher forms in $\Phi$, including even the original action. For the reasons we discuss next, the simplest way to read off $\Phi_0$ turns out to be the direct evaluation of the action on the Omega background in AdS$_4$ exhibiting precisely one fixed point.

\section{Off-shell supergravity,  CKS backgrounds and\\ higher derivative invariants}
\label{sec:sphere}

We now restrict our attention to 4d $\cN=2$ supergravity, which in itself is a vast subject with a long history, see e.g.\ \cite{Lauria:2020rhc,Kuzenko:2022ajd} for recent reviews. In particular we consider conformal supergravity, which is a completely off-shell formulation of the theory with an extended set of local symmetries that facilitate the construction of supersymmetric Lagrangians with any number of derivatives and considerable freedom in the choice of matter couplings, see \cite{Ozkan:2024euj}. 

We directly start with the one indispensable supergravity multiplet, which contains the metric and the gravitini.~\footnote{In the superconformal formalism, we already have the choice of using either the so-called standard Weyl multiplet, or the dilaton Weyl one. We choose the former option, in line with the choice in \cite{Hristov:2021qsw}, which also dictates the way supersymmetric invariants are built.} The contents of this miltiplet are the gauge fields for the local symmetries of the theory, together with supersymmetric completions. The bosonic symmetry algebra (in Euclidean signature, see \cite{Klare:2013dka,deWit:2017cle}) includes the general coordinate, local Lorentz, dilatation and special conformal transformations, as well as $SO(1,1) \times SU(2)$ R-symmetries, while the fermionic symmetries are divided into supersymmetry (Q) and conformal supersymmetry (S) transformations. We thus have the vielbein $e_\mu^a$, the $SU(2)$ gauge field $\cV_\mu{}^i{}_j$, the conformal and $SO(1,1)$ gauge fields $b_\mu, A_\mu$, as well as two gravitini $\psi^i_\mu$. The additional fields, which in the two derivative theory are auxiliary, include the antisymmetric tensor $T_{\mu\nu}^\pm$ (the $\pm$ parts are independent in Euclidean signature), a scalar D and two spin-$1/2$ fermions $\chi^i$. Importantly, all fermions (including the $Q$ and $S$ variation parameters used below, $\varepsilon^i_\pm$ and $\eta^i_\pm$, respectively) can be split into independent chiral ($+$) and anti-chiral ($-$) pieces.

{\bf BPS backgrounds.} As discussed in the previous sections at a very general level, we are interested in BPS backgrounds, which preserve some amount of supersymmetry. Although with a different motivation, the complete classification of such backgrounds has already been addressed and accomplished in \cite{Klare:2013dka} (assuming as usual vanishing fermions). Let us focus on the gravitino variation,
\be
	 \delta \psi_\mu^i{}_\pm = D_\mu \epsilon^i_\pm + \frac{\i}{4} T^{\mp}_{\mu\nu} \gamma^\nu \epsilon^i_\mp -\i\, \gamma_\mu \eta^i_\mp\ ,
\ee
where
\be
	D_\mu \epsilon^i_\pm := (\partial_\mu - \tfrac14 \omega_\mu{}^{ab} \gamma_{ab})\, \epsilon^i_\pm + (b_\mu \pm A_\mu)\, \epsilon_\pm + \cV_\mu{}^i{}_j \epsilon^j_\pm\ .
\ee
Defining $\slashed{D} = \gamma^\mu D_\mu$, the vanishing of the gravitino variation requires $\eta_\pm^i = \tfrac1{4 i} \slashed{D} \epsilon_\mp^i$, such that
\be
\label{eq:KSI}
	 D_\mu \epsilon^i_\pm + \frac{\i}{4} T^{\mp}_{\mu\nu} \gamma^\nu \epsilon^i_\mp - \tfrac14 \gamma_\mu \slashed{D} \epsilon_\pm^i = 0\ .
\ee
This is the equation for a (generalized) conformal Killing spinor, abbreviated as CKS. Assuming a solution to the above equation, it is a standard procedure to define a set of conformal Killing spinor bilinears, most notably the vector
\be
	\xi^\mu := \frac{i}{2}\, \epsilon^\dagger_{i, -} \gamma^\mu \epsilon^i_+\ ,
\ee
and from \eqref{eq:KSI} it can be shown that
\be
	\nabla_\mu \xi_\nu + \nabla_\nu \xi_\mu = \lambda\, g_{\mu \nu}\ ,
\ee
i.e.\ the vector $\xi$ is a conformal Killing vector, or CKV,
\be
	\cL_\xi\, g = \lambda\, g\ ,
\ee
for some function $\lambda$. What is important for the present purposes, and again proven in \cite{Klare:2013dka} in analogy to \cite{Klare:2012gn}, is that the \emph{converse} implication is also true: the existence of a CKV is a necessary and sufficient condition for the existence of a CKS. The only geometric constraint on BPS backgrounds is therefore the existence of a single conformal isometry. For the purposes of equivariant localization, the only additional assumption we need is that the conformal isometry is \emph{compact}. Furthermore, it is clear that conformally equivalent backgrounds are practically indistinguishable in this formalism, such that we can always look at the Weyl representative, for which $\xi$ is an exact isometry. This was utilized in \cite{Klare:2013dka} to constrain the form of the metric, but is not needed from the present perspective.

Note that there are other constraints coming from supersymmetry, for example the tensor $T^\pm_{\mu\nu}$ is determined by solving the dilatino variation $\delta \chi^i_\pm = 0$. Additionally, we need to make sure all fermionic variations from matter multiplets vanish as well. An important feature of off-shell supersymmetry is that the supersymmetry variations of different multiplets are \emph{decoupled} from each other. We are thus free to add an arbitrary set of additional matter multiplets without imposing any additional conditions on the geometry, the $T$-tensor or any other Weyl multiplet fields. In the language of equivariant localization it means matter multiplets do not affect the identification of equivariant parameters on a given background. Instead, the extra BPS conditions on the matter fields enter in the gluing rules via what we called Coulomb branch parameters.

{\bf Matter content and higher derivative invariants.} There are various matter multiplets, stemming from corresponding superfields, which can either represent new fundamental fields or composite objects with the aim of building supersymmetric invariants. An important example is the usual vector multiplet, which can alternatively be viewed as a reducible combination of a chiral and anti-chiral scalar multiplet, \cite{deWit:2017cle}. We consider $(n_V+1)$ vector multiplets, $I = 0, ..., n_V$ with field content
\be
	\cX^I = \{X^I_\pm, \Omega^{I, i}_\pm, W^I_\mu, Y^I_{ij} \}\ ,
\ee
where $W^I_\mu$ are vectors (here abelian by choice), $\Omega^{I, i}$ spin-$1/2$ fermions, $X^I_\pm$ pairs of real scalars (a single complex scalar $X^I$ in Lorentzian signature), and $Y^I_{ij}$ triplets of real scalars. We include also a coupling to a single (auxiliary for the on-shell theory) hypermultiplet with four real scalars and two spin-$1/2$ fermions, which allows for the \emph{gauging} of the R-symmetry with some linear combination of the vectors. For an on-shell theory with AdS$_4$ vacuum (i.e.\ a holographic theory), we need to break the $SU(2)$ R-symmetry to a $U(1)_R$ and pick a set of constants $g_I$ that embed the corresponding linear combination of gauge fields $W^I$ into it. The resulting theory is called Fayet-Iliopoulos (FI) gauged supergravity, and naturally admits a limit to the ungauged version when $g_I \rightarrow 0$. This is not a unique choice as there are alternative ways of introducing the gauging, e.g.\ using the so called linear multiplet, see \cite{Gold:2023ymc} for the analogous 5d construction.
 
The formalism also allows a considerable freedom when building higher derivative Lagrangians. At four derivatives there are three invariants that can be regarded as the supersymmetric completions of three different curvature squared terms: the Weyl$^2$ action, \cite{Bergshoeff:1980is}, the logarithm of the kinetic multiplet action including the Gauss-Bonnet invariant, \cite{Butter:2013lta}, and the $R^2$ invariant, \cite{deWit:2006gn,Butter:2010jm}. All these actions can in turn be used as building blocks for further invariants of higher orders, which are classified as $F$-terms, or chiral superspace integrals. There are also full superspace invariants, or $D$-terms, such as \cite{deWit:2010za}, which are expected to vanish on BPS backgrounds.

Following \cite{Hristov:2021qsw}, we focus on the higher derivative theory with the following fundamental multiplets: the Weyl multiplet, $(n_V + 1)$ vector multiplets and one hypermultiplet with FI parameters $g_I$ as described above.~\footnote{Note that strictly speaking the choice of using a hypermultiplet to provide the gauging is \emph{not} consistent in the presence of higher derivatives. This is due to the inherent inability for the hypermultiplet to be fully off-shell. One should instead use linear multiplets to include the gauging. This technical point is not crucial for the present arguments, which are not based on the actual Lagrangian, so the results of \cite{Bobev:2020egg,Bobev:2021oku,Hristov:2022plc} are not expected to be sensitive to this issue. We thank Gabriele Tartaglino-Mazzucchelli for discussions on this point.} We include only two of the (at least) three possible building blocks for higher derivative invariants: the Weyl$^2$ action with a lowest component the composite scalar $A_\mathbb{W}$, and the T-$\log$ action with lowest component the composite scalar $A_\mathbb{T}$ (see \cite{Hristov:2021qsw,Hristov:2022plc} for the precise form of the action).~\footnote{We stress that the existence of other off-shell higher derivative invariants does not mean they result in independent on-shell actions. Thus, even though the third building block is omitted, we expect that it does not give linearly independent results.} The full higher derivative action then takes the form of an infinite expansion in powers of these main building blocks, such that it can be uniquely defined by the following function, called the prepotential
\be
\label{eq:1}
	F (X^I; A_\mathbb{W}, A_\mathbb{T}) = \sum_{m, n = 0}^\infty F^{(m,n)} (X^I)\, (A_\mathbb{W})^m\, (A_\mathbb{T})^n\ ,  
\ee
with gauging parameters $g_I$ and the extra assumptions (for compatibility with a Lorentzian signature version) $X^I_+ = X^I_-$. The functions $F^{(m,n)} (X^I)$ are always homogeneous and have a fixed degree in terms of the scalars $X^I$, given by $2\, (1 - m - n)$, and respectively give rise to terms with $2 (1+m+n)$ derivatives. Finally, we should stress that in the superconformal formalism there are a priori no scales, and only the choices for gauge fixing break the (super) conformal symmetries. The explicit appearance of the Newton constant is thus a particular choice. We implicitly keep the logic that each higher derivative term is further suppressed, but explicitly only rescale the full action by an overall negative power of $G_N$ while assuming any other powers of $G_N$ are hidden within the definitions of the functions $F^{(m,n)}$.

\section{Omega background and proof of \cite{Hristov:2021qsw}}
\label{sec:omega}
We now turn to a brief summary of \cite{Hristov:2022plc}, where the Omega backgrounds in supergravity were considered in detail. More specifically, it was shown that the gauged supergravity generalization of the usual $\Omega \mathbb{R}^4$ background of Nekrasov-Okounkov, \cite{Nekrasov:2003rj}, was already discovered in literature as the supergravity dual of the squashed three-sphere, \cite{Martelli:2011fu}. Geometrically, the background is simply the maximally symmetric vacuum, Euclidean AdS$_4$ or simply $\mathbb{H}^4$. However, we explicitly break the full symmetry group down to $U(1) \times U(1)$ by adding either a purely self-dual or a purely anti-self-dual tensor field $T_{\mu \nu}$. Without loss of generality we can pick $T^+ =0$ as in \cite{Hristov:2022plc}. Since we additionally include $(n_V + 1)$ vectors and a hypermultiplet introducing the gauging parameters $g_I$, we find an additional set of BPS constraints on these fields. 

The full set of background fields is presented in \cite{Hristov:2022plc}, and here we reproduce only the formulae that have a direct importance for equivariant localization. The metric of  $\Omega\, \mathbb{H}^4$, for simplicity working with a unit radius $L=1$, is
\be
	{\rm d} s^2 = \frac{{\rm d} r^2}{1+r^2} + r^2 \left( {\rm d} \theta^2 + \cos^2 \theta\, {\rm d} \varphi_1^2 + \sin^2 \theta\, {\rm d} \varphi_2^2 \right) \ ,	
\ee
which is the hyperbolic space $\mathbb{H}^4$, radially sliced in round S$^3$ coordinates. The $T$-tensor in these coordinates reads
\be
\label{eq:Tform}
\begin{split}
	   T^+_{\mu\nu} & = 0\ , \quad  T^-_{\theta \varphi_1} = - \frac{2 (b_1^2-b_2^2) (b_1 + b_2 \sqrt{1+r^2})\, r^2 \sin \theta \cos \theta}{ \Xi^{3/2} (r, \theta)}= \frac{ (b_1 + b_2 \sqrt{1+r^2}) }{(b_2 + b_1 \sqrt{1+r^2})}  T^-_{\theta \varphi_2} \ , \\
T^-_{r \theta} & = 0\ , \quad T^-_{r \varphi_1} =  \frac{2 (b_1^2-b_2^2) (b_2 + b_1 \sqrt{1+r^2})\, r \cos^2 \theta}{\sqrt{1+r^2}\, \Xi^{3/2} (r, \theta)}= - \frac{ (b_2 + b_1 \sqrt{1+r^2}) \cos^2 \theta}{(b_1 + b_2 \sqrt{1+r^2}) \sin^2 \theta}  T^-_{r \varphi_2} \ ,
\end{split}
\ee
where
\be
	\Xi(r, \theta) :=(b_1 + b_2 \sqrt{1+r^2})^2 \sin^2 \theta +  (b_2 + b_1 \sqrt{1+r^2})^2 \cos^2 \theta\ .
\ee

In the center of the space, $r \rightarrow 0$, we find the usual locally flat Omega background in spherical Hopf coordinates, 
\be
		\lim_{r \rightarrow 0}\, {\rm d} s^2 = {\rm d} r^2 + r^2 \left( {\rm d} \theta^2 + \cos^2 \theta\, {\rm d} \varphi_1^2 + \sin^2 \theta\, {\rm d} \varphi_2^2 \right) \ ,	
\ee
with non-vanishing anit-self-dual $T^-$ following from \eqref{eq:Tform}.
Here we can identify directly the equivariant parameters (see \cite{Hristov:2022plc} for details), $b_1$ and $b_2$ responcible for the rotations around the two $U(1)$'s. In particular, we find the Killing spinor bilinear
\be
	\xi = b_1\, \partial_{\varphi_1} + b_2\, \partial_{\varphi_2}\ ,
\ee
upto an arbitrary and irrelevant overall scale. Note that $\xi$ is automatically an exact isometry of the background, not just a conformal one. It is clear that the center of the space is the unique fixed point of this $U(1)$ action, as already expected for an Omega background. As already stressed, the equivariant parameters $b_1$ and $b_2$ can be identified just by looking at the Weyl multiplet.

The important BPS conditions for us stemming from the gaugini and hyperino variations are the following. The scalar fields $X^I_+ = X^I_- = X^I$ are all set to constants, which are further constrained as
\be
	g_I X^I = 1\ ,
\ee
in inverse units of the AdS scale. The gauge fields and their respective field strengths are then fully determined by the $T$-tensor,
\be
	 F^{I, \pm}_{\mu\nu} = \frac14 X^I T^\pm_{\mu\nu}\ ,
\ee
leading automatically to purely anti-self-dual field strengths, $F^{I +}_{\mu\nu} = 0$. This makes it evident that the only remaining unfixed parameters remain the scalars $X^I$. It is precisely the $X^I$ that can be thought of as Coulomb branch parameters, and the above constraint is a particular rule for them that follows from supersymmetry.

{\bf Proof of \cite{Hristov:2021qsw}.} Starting from the full higher derivative action, defined in \eqref{eq:1}, see Eq. (3.3) in \cite{Hristov:2022plc}, it was then shown that the $\Omega\, \mathbb{H}^4$ background enjoys a particularly nice property for the purposes of holographic renormalization. Simiilarly to many examples in \cite{Bobev:2021oku}, the diverging pieces of the action on this background turn out to be canceled by the standard two derivative counterterms that include the Gibbons-Hawking-York action, \cite{Gibbons:1976ue,York:1986it}. Explicitly, it was shown that the following boundary terms render the action finite,
\be
	I_\text{bdry} = - \int_{\partial M} {\rm d} x^3 \sqrt{h}\, F(2 X^I; 0, 4)  \left( K - \tfrac12\, \cR - 2 \right)\ , 
\ee
where $h$ is the induced metric on the boundary, $K$ the extrinsic curvature, and $\cR$ is the Ricci scalar of the induced metric. Notice that we are not claiming these are all terms needed for holographic renormalization of an \emph{arbitrary} background, but only specifically the ones relevant for $\Omega\, \mathbb{H}^4$. Therefore, in the language of equivariant localizations we cannot present the general form of $\Phi_4$ (c.f.\ \eqref{eq:preequiloc}), but we can still evaluate on the Omega background. After a somewhat technical, but in principle straightforward calculation, the final action is directly evaluated to be, \cite{Hristov:2022plc},
\be
	I (\Omega\, \mathbb{H}^4) (b_1, b_2; X^I)= \frac{4 \pi^2}{\kappa^2\, b_1 b_2} F\left((b_1+b_2) X^I; (b_1-b_2)^2,  (b_1 + b_2)^2\right) \ .
\ee

Since we have a single fixed point, we can read off the zero-form in \eqref{eq:multiform} directly from \eqref{eq:equiloc},
\be
	\Phi_0 (b_1, b_2; X^I) = \frac{4 \pi^2}{\kappa^2} F\left((b_1+b_2) X^I; (b_1-b_2)^2,  (b_1 + b_2)^2\right) \ ,
\ee
such that the off-shell action for BPS backgrounds with fixed points is given by
\be
\label{eq:final}
	I (M) = \sum_{\sigma}\, \frac{\Phi_0  (b^{(\sigma)}_1, b^{(\sigma)}_2; X^I_{(\sigma)})}{b^{(\sigma)}_1\, b^{(\sigma)}_2}\ .
\ee
Observe that the expression at a single fixed point is manifestly dependent only on the ratio $\omega := b_2 / b_1$ and is also invariant under $\omega \rightarrow 1/\omega$,
\be
\label{eq:final-alt}
	I (M) =  \frac{4 \pi^2}{\kappa^2}\sum_{\sigma}\, \frac{F\left((1+\omega_{(\sigma)}) X^I_{(\sigma)}; (1-\omega_{(\sigma)})^2,  (1 + \omega_{(\sigma)})^2\right) }{\omega_{(\sigma)}}\ ,
\ee
due to the simple homogeneity property of the prepotential, \eqref{eq:1}. This is the essence of the main conjecture, presented in \cite{Hristov:2021qsw} (part I).~\footnote{Note that the parameters $X^I$ in  \cite{Hristov:2021qsw} are rescaled, $ X^I_\text{there} = (b_1 + b_2)\, X^I_\text{here}$, such that the Coulomb branch constraint $g_I X^I_\text{here} = 1$ is also rescaled accordingly. This constraint is also part of the gluing rules, and thus changes with the respective background.} It is remarkable that the equivariant parameters $b_{1,2}$, or $\omega$, are solely responsible for the full higher derivative expansion. Special values of $\omega$ are thus associated with the so-called unrefined and Nekrasov-Shatashvili limits, as well as the Cardy regime, see \cite{Hristov:2021qsw} for details.

Finally, it was also shown in \cite{Hristov:2022plc}, in accordance with \cite{Hristov:2021qsw}, that solving the equations of motion for the scalar fields is equivalent with extremization of the off-shell action,~\footnote{Although so far we have \emph{not} used any equations of motion, it is a standard feature of supersymmetric backgrounds that they automatically satisfy some of the underlying field equations, see \cite{Kallosh:1993wx}. The only remaining equations in this case are the scalar ones.}
\be
	I^\text{on-shell} (\Omega\, \mathbb{H}^4) (b_1, b_2) =  I (\Omega\, \mathbb{H}^4) (b_1, b_2; \overline X^I)\ , \qquad  \frac{\partial I (\Omega\, \mathbb{H}^4) (b_1, b_2; \overline X^I) }{\partial X^I} \Big|_{\overline X^I} = 0 \ .
\ee
We expect the above principle to hold more generally for any supersymmetric background $M$, but this can only be proven on a case by case basis as it does not immediately relate to the equivariant localization theorem.

\section{Gluing rules and black holes}
\label{sec:setup}
We have already emphasized in several places above that the off-shell formalism significantly simplifies the concept of gluing rules for BPS backgrounds with fixed points under the $U(1)$ action $\xi$. To summarize the discussion, we distinguish between two types of scalar parameters in the final form of the action, \eqref{eq:final}:
\begin{itemize}
	\item {\bf Equivariant parameters, $b^{(\sigma)}_1, b^{(\sigma)}_2$.}Together with the explicit position and number of fixed points $\sigma$, the equivariant parameters can be identified from the background form of the Weyl multiplet, in particular from the local form of two-form field $T^\pm$ in the neighborhood of the fixed point.

	\item {\bf Coulomb branch parameters, $X^I_{(\sigma)}$.} These are all additional parameters coming from matter couplings, in the explicit case that we consider the scalars in vector multiplets.
\end{itemize}
Importantly, none of the above depends on the choice of the Lagrangian. This means that the gluing rules for a general higher derivative supergravity are precisely the same as in the better studied two derivative Poincar\'e theory, as long as the underlying physical matter matches. Thus we can directly use the gluing rules for various 4d black hole spacetimes, discussed and derived from different perspectives in \cite{Hosseini:2019iad,Boido:2022iye,Boido:2022mbe,BenettiGenolini:2024kyy}, which are prime examples of backgrounds with fixed points.

{\bf Black hole rules.} For concreteness, let us reproduce the black hole gluing rules in the present conventions.~\footnote{Since we only rely on Euclidean supersymmetry, we do not distinguish here between spaces that correspond (upon Wick rotation) to Lorentzian black holes and Euclidean black saddles discussed in 4d in \cite{Cassani:2019mms,Bobev:2020pjk,Hristov:2022pmo}. These are indeed topologically indistinguishable in equivariant localization and produce the same off-shell results. The distinction arises only if equations of motions are imposed and depends on the reality conditions on the matter multiplets.} We can directly consider the general case of spindle topology of the horizon, see \cite{Ferrero:2020twa}, and look at both ways of preserving supersymmetry, dubbed twist and anti-twist, \cite{Ferrero:2021etw}.~\footnote{At the level of the near-horizon geometry, all black hole solutions in abelian gauged two derivative supergravity were written down in \cite{Hristov:2018spe} for the twist case and \cite{Hristov:2019mqp} for the anti-twist case in presence of general electromagnetic charges and rotation.}

Focusing directly at the near-horizon regions of these black holes, which is where the fixed points are, we consider the space $\mathbb{H}^2 \times \mathbb{WP}^1_{n_-, n_+}$. The former factor is the Euclidean AdS$_2$ space typically appearing for extremal black holes, while the latter factor is colloquially known as the spindle with two co-prime numbers $n_\mp$. characterizing its conical singularities. Following the split of the geometry, the corresponding $U(1)$ action, $\xi$, can also be considered on the two spaces separately. On $\mathbb{H}^2$, which has the same topology as the complex plane $\mathbb{C}$, it acts as the usual rotation leaving only the origin fixed. On the spindle, $\xi$ acts locally as a rotation around the (unique) symmetry axis, leaving fixed the two poles. We thus find two fixed points on the full 4d space, corresponding to the product of the center of $\mathbb{H}^2$ with the two poles. Taking in consideration the fluxes $p^I$ through the spindle, the gluing rules at these two points, $\sigma = \{\text{NP}, \text{SP} \}$ are the given by:
\begin{itemize}
	\item {\bf Equivariant parameters, $b^{(\sigma)}_1, b^{(\sigma)}_2$.} As already observed, the building blocks only depend on the ratio of $b_1$ and $b_2$, which allows us for a more straightforward identification with the geometric parameter $\omega$ of the black holes,
\be
	b_2 / b_1 |_\text{NP} = \omega\ , \qquad b_2 / b_1 |_\text{SP} = s\, \omega\ , 
\ee
where $s = \pm 1$ is the important sign that distinguishes between the twist, $s = -1$, and the anti-twist, $s = +1$.

	\item {\bf Coulomb branch parameters, $X^I_{(\sigma)}$.}~\footnote{Here we only consider a single auxiliary hypermultiplet, in agreement with the matter content in \cite{Hristov:2021qsw}. Additional physicsl hypermultiplets typically lead to massive vectors and further supersymmetric constraints, see \cite{Hristov:2023rel,Hristov:2024qiy} for such models.} We make use of the rescaling,
\be
	\bar X^I_{(\sigma)} := (1 + \omega_{(\sigma)}) X^I_{(\sigma)}\ ,
\ee
to arrive at the simpler gluing rules
\be
	\bar X^I |_\text{NP} = \chi^I - \omega\, p^I\ , \qquad \bar X^I |_\text{SP} = \chi^I + \omega\, p^I\ , 
\ee
where the scalar parameters and the magnetic fluxes are constrained by supersymmetry,
\be
\label{eq:constr}
	g_I \chi^I = 1 + \frac{n_+ + s\, n_-}{2 n_+ n_-}\, \omega\ , \qquad g_I p^I = \frac{n_+ - s\, n_-}{2 n_+ n_-}\ .
\ee
\end{itemize}

Taking all these identifications into account, we arrive at the higher derivative action of black holes,
\be
	I (\omega, \chi^I) =  \frac{4 \pi^2}{\kappa^2\, \omega}\, \left( F (\chi^I - \omega\, p^I; (1 - \omega)^2, (1+\omega)^2) + s\, F (\chi^I + \omega\, p^I; (1 - s\, \omega)^2, (1+s\, \omega)^2) \right)\ ,
\ee
under the constraints \eqref{eq:constr}. Importantly, the parameters $\chi^I$ are precisely the chemical potentials conjugate to the electric charges of the black hole, while $\omega$ is conjugate to the angular momentum, allowing to define the black hole entropy function as a Legendre transform of the action, see \cite{Hristov:2021qsw}. The above set of equations generalizes the black hole gluing rules in \cite{Hristov:2021qsw} by admitting arbitrary values of the spindle parameters, $n_\pm$.

Finally note that, as already discussed at length in \cite{Hristov:2021qsw}, the black hole action and entropy function are also well defined in the ungauged limit, $g_I = 0$. In this case supersymmetry fixes the spherical condition, $n_+ = n_- = 1$ above, and thus only allows for the anti-twist conditions, $s = +1$ and $\omega = -1$, as otherwise \eqref{eq:constr} cannot be solved.~\footnote{Note also that the above gluing rules are naturally adapted from the two derivative description of asymptotically AdS black holes, as in \cite{Hosseini:2019iad}, with electric gauging parameters $g_I$. Depending on the frame, in which the prepotential is defined, there could be an additional sign in the gluing rules that becomes manifest in the ungauged limit, see \cite{Hristov:2022pmo}. This technical detail is more naturally addressed at the level of the complete internal geometry inside string theory, \cite{Martelli:2023oqk}.}

\section{M2-brane partition functions}
\label{sec:m2}
Higher derivative supergravity can be used in holographic settings to predict or match dual field theory observables. An older example is related to black holes in ungauged supergravity, see \cite{LopesCardoso:1998tkj,Ooguri:2004zv}, where the $\mathbb{T}$-tower of higher derivative invariants vanishes (due to $\omega = -1$). A more recent application concerns M2-brane partition functions, starting from the four derivative results in \cite{Bobev:2020egg,Bobev:2021oku}. Based on the full infinite derivative expansion of BPS backgrounds in \eqref{eq:final-alt} and the localization results of \cite{Fuji:2011km,Marino:2011eh,Nosaka:2015iiw,Hatsuda:2016uqa,Chester:2021gdw}, \cite{Hristov:2022lcw} formulated a very precise prediction for the dual ABJM partition functions at finite gauge group rank $N$. Here we point out that the same logic can be similarly applied to other M2-brane models, not necessarily put on the tip of $\mathbb{C}^4 / \mathbb{Z}_k$ but in many other supersymmetric settings.

The main idea is that the Omega-AdS$_4$ background should holographically match with the respective squashed sphere partition function as a full perturbative expansion in the respective coupling constants. This might sound somewhat abstract, but there are general arguments, \cite{Marino:2011eh}, why the squashed S$^3_\omega$ partition functions M2-brane models always takes the form of an Airy function, see \cite{Geukens:2024zmt} and references thereof for more recent results. This already gives us a handle on holography in a very practical form. Let us look at the asymptotic form of the Airy function, which would be valid in a large $N$ expansion that matches an effective supergravity description,
\be
\label{eq:asymp}
Z_{S^3_\omega} = \text{Ai} (z) \sim \frac{e^{-2/3\, z^{3/2}}}{2 \sqrt{\pi} z^{1/4}}\, \Big[ \sum_{n=0}^\infty \frac{(-1)^n 3^n \Gamma(n+\frac56) \Gamma(n+\frac16)}{2 \pi n! 4^n z^{3n/2} } \Big]  \ ,
\ee
where $z$ represents a collection of field theory variables and parameters. In all known examples of M2-brane models, corresponding to 3d Chern-Simons-matter theories with a gauge group rank $N$, $z$  is given by a linear and a constant term in $N$, both of which depend on the (complexified) R-symmetry charges $\Delta_i$ and the geometric squashing parameter $\omega$,
\be
	z = N\, f_1 (\omega, \Delta_i) + f_2 (\omega, \Delta_i)\ .
\ee
 Irrespective of the specific details, the obvious consequence of the Airy function expansion is that the exponential can be very clearly isolated,
\be
	\log Z_{S^3_\omega} \approx -\tfrac23\, z^{3/2} - \tfrac14\, \log z + ... =  -\tfrac23\, \left( N\, f_1 (\omega, \Delta_i) + f_2 (\omega, \Delta_i) \right)^{3/2} + ... \ ,
\ee
where we neglected further subleading terms in $N$. The main point is that effective supergravity presicely reproduces the exponential behavior of the full partition function, since we are just evaluating the exponent of the classical action from a full path integral point of view. Further logarithmic and higher order corrections do not appear from the direct evaluation of the supergravity action.~\footnote{One might hope to partially understand logarithmic correction in supergravity via the Atiyah-Singer index theorem \cite{atiyah1963index}, see \cite{Hristov:2021zai,Bobev:2023dwx}, but ultimately it is clear that all corrections not manifest in the effective action must be analysed directly in string theory.} 
Explicitly, this means the holographic equation
\bea
\label{eq:abjmequation}
\begin{split}
	-\log Z_{S^3_\omega}  \approx& \tfrac23\, \left( N\, f_1 (\omega, \Delta_i) + f_2 (\omega, \Delta_i) \right)^{3/2} \\
	&\stackrel{!}{=}	I (\Omega\, \mathbb{H}^4) =  \frac{4 \pi^2}{\kappa^2\, \omega}\, F\left((1+\omega) X^I; (1-\omega)^2,  (1 + \omega)^2\right)\ ,
\end{split}
\eea
where the R-charge parameters $\Delta_i$ are identified (upto rescaling) with the Coulomb branch parameters $X^I$, based on the asymptotic symmetries, \cite{Benini:2015eyy}. If we Taylor expand the field theory answer at large $N$, we find an infinite series in $N$ that needs to match to an infinite series in higher derivatives, i.e.\ powers of $\kappa^2$. 

Remarkably, this equation can be solved \emph{exactly} for the ABJM model, determining the higher derivative coefficients $F^{(m,n)}$ in \eqref{eq:1} order by order, using a number of constraints and results on both sides of the duality, \cite{Hristov:2022lcw}. Once the single gravitational block is determined in this holographic fashion, using the general form of \eqref{eq:final-alt} clearly allows to determine many other partition functions. Such a successful holographic match was already performed for a subset of the black hole solutions with a twist from the previous section, corresponding to the topologically twisted index, \cite{Benini:2015eyy,Bobev:2022eus}.

{\bf Airy conjecture.} Even more remarkably, observe that once we have identified correctly the exponent $z$ in \eqref{eq:asymp}, we can reproduce the \emph{entire} Airy function. Therefore, even if the quantum corrections are only well-defined in string theory, supergravity allows us to formulate an obvious conjecture for M2-brane partition functions,  \cite{Hristov:2022lcw}: every fixed point of the gravitational background contributes precisely with a factor of an Airy function, the respective argument being determined from the gluing rules. It would be interesting to verify this conjecture directly in string theory again based on equivariant localization, see the discussion in the next section. 

An important remark related to this point is that the form of the gravitational building block in \eqref{eq:final-alt} as a Nekrasov-like partition function resembles a \emph{refined} topological string expansion by the fact that there are two independent towers of corrections stemming from the $\mathbb{W}$ and $\mathbb{T}$ invariants, and this appears to capture the dual ABJM partition function given the explicit solution of \eqref{eq:abjmequation} in \cite{Hristov:2022lcw}.

\section{List of open questions}
\label{sec:sci}
There are some closely related questions that can be grouped together in many meaningful ways. One possibility is the following list.
\begin{itemize}

\item {\bf Omega backgrounds in $d > 4$.}\\
The primary aim and contribution of this note is to prove the conjecture of \cite{Hristov:2021qsw}. We achieved this by a shortcut: selecting the particularly simple Omega background, with only a single fixed point under the $U(1)$ action used for equivariant localization. By directly evaluating the supergravity action, we found the bottom-form $\Phi_0$ and bypassed the complexities of defining the action on an arbitrary background and computing its equivariant completion. This significant simplification could be very beneficial in other dimensions, such as $d=6,8,10$.~\footnote{Some equivariant localization results for supergravity actions in these dimensions can already be found in \cite{BenettiGenolini:2023kxp,BenettiGenolini:2023ndb,Couzens:2024vbn}.} Constructing the relevant Omega backgrounds and evaluating the supergravity actions should lead to similarly general expressions like \eqref{eq:final-alt} for large classes of BPS backgrounds. \\

\item {\bf Equivariant completion of the action.}\\
As discussed above, here we bypassed the equivariantly closed completion of the action. However, performing this calculation is very interesting in itself, as it could provide insights into the fundamental meaning of holographic renormalization and its rigorous mathematical formulation. Additionally, formulating the supergravity action as an equivariantly closed form would be key to extending rigorously our results to backgrounds with fixed two-submanifolds. We hope to report on this in future work \cite{Hristov-Hosseini-Reys}.\\

\item {\bf 5d higher derivative supergravity.}\\
The method of equivariant localization can significantly simplify the action of even-dimensional backgrounds, whether they are compact or non-compact (with appropriate boundary terms). See \cite{Martelli:2023oqk,BenettiGenolini:2023ndb,BenettiGenolini:2024kyy,Suh:2024asy} for related results. Although the five-dimensional, K-theoretic, analog of Nekrasov's Omega background is well-understood in field theory, it remains challenging to replicate the same steps in supergravity. On the other hand, the superconformal formalism has been extensively developed for five-dimensional supergravity (see \cite{Gold:2023ymc, Gold:2023ykx}), where the construction of higher derivative invariants shares many similarities with the 4d case. Therefore, it is natural to expect that the general form of the action on BPS backgrounds can be expressed in a single formula, to be discussed in \cite{Hristov-Yi-Gab}.\\

\item {\bf Equivariant localization in string theory.}\\
Given that effective supergravity is only the low-energy limit of string theory, which provides the full UV completion, it makes sense to look directly in the parent theory. This was already initiated in \cite{Martelli:2023oqk} by calculating the equivariant generalization of the symplectic volume of toric manifolds in string theory models. From the perspective of topological string theory, the equivariant generalization of Calabi-Yau intersection numbers can be viewed as a natural extension. This appears to be the appropriate framework to derive quantum gravity results such as the Airy conjecture in \cite{Hristov:2022lcw}, and to extend it non-perturbatively. This will be discussed in \cite{Hristov-Cassia}.\\

\item {\bf Non-supersymmetric extension.}\\
The simplification of the supergravity action, based on equivariant localization, technically relies only on a $U(1)$ action. This feature is also present in interesting non-BPS backgrounds, such as thermal black holes, which exhibit a rather different form of the supergravity action, see \cite{Hristov:2023sxg,Hristov:2023cuo}.. It would be intriguing to formulate an equivariant localization calculation for these backgrounds, potentially unveiling a general principle for evaluating higher derivative (and further quantum and string theoretic) corrections for thermal black holes.
\end{itemize}

\section*{Acknowledgements}
I am very grateful to Luca Cassia, Seyed Morteza Hosseini, Yi Pang, Valentin Reys, and Gabriele Tartaglino-Mazzucchelli for useful discussions and inspiring collaborations on related topics. I am also indebted to the organizers and participants of the MATRIX Program “New Deformations of Quantum Field and Gravity Theories,” between 22 Jan and 2 Feb 2024 for the kind hospitality. This study is financed by the European Union- NextGenerationEU, through the National Recovery and Resilience Plan of the Republic of Bulgaria, project No BG-RRP-2.004-0008-C01. 

\bibliographystyle{spmpsci}
\bibliography{matrix.bib}

\begin{thebibliography}{10}
\providecommand{\url}[1]{{#1}}
\providecommand{\urlprefix}{URL }
\expandafter\ifx\csname urlstyle\endcsname\relax
  \providecommand{\doi}[1]{DOI~\discretionary{}{}{}#1}\else
  \providecommand{\doi}{DOI~\discretionary{}{}{}\begingroup
  \urlstyle{rm}\Url}\fi

\bibitem{Atiyah:1984px}
Atiyah, M.F., Bott, R.: {The Moment map and equivariant cohomology}.
\newblock Topology \textbf{23}, 1--28 (1984).
\newblock \doi{10.1016/0040-9383(84)90021-1}

\bibitem{atiyah1963index}
Atiyah, M.F., Singer, I.M.: The index of elliptic operators on compact
  manifolds.
\newblock Bulletin of the American Mathematical Society \textbf{69}(3),
  422--433 (1963)

\bibitem{BenettiGenolini:2024kyy}
Benetti~Genolini, P., Gauntlett, J.P., Jiao, Y., L\"uscher, A., Sparks, J.:
  {Localization and attraction}.
\newblock JHEP \textbf{05}, 152 (2024).
\newblock \doi{10.1007/JHEP05(2024)152}

\bibitem{BenettiGenolini:2023kxp}
Benetti~Genolini, P., Gauntlett, J.P., Sparks, J.: {Equivariant Localization in
  Supergravity}.
\newblock Phys. Rev. Lett. \textbf{131}(12), 121,602 (2023).
\newblock \doi{10.1103/PhysRevLett.131.121602}

\bibitem{BenettiGenolini:2023ndb}
Benetti~Genolini, P., Gauntlett, J.P., Sparks, J.: {Equivariant localization
  for AdS/CFT}.
\newblock JHEP \textbf{02}, 015 (2024).
\newblock \doi{10.1007/JHEP02(2024)015}

\bibitem{BenettiGenolini:2019jdz}
Benetti~Genolini, P., Perez Ipi\~na, J.M., Sparks, J.: {Localization of the
  action in AdS/CFT}.
\newblock JHEP \textbf{10}, 252 (2019).
\newblock \doi{10.1007/JHEP10(2019)252}

\bibitem{Benini:2015eyy}
Benini, F., Hristov, K., Zaffaroni, A.: {Black hole microstates in AdS$_{4}$
  from supersymmetric localization}.
\newblock JHEP \textbf{05}, 054 (2016).
\newblock \doi{10.1007/JHEP05(2016)054}

\bibitem{Bergshoeff:1980is}
Bergshoeff, E., de~Roo, M., de~Wit, B.: {Extended Conformal Supergravity}.
\newblock Nucl. Phys. B \textbf{182}, 173--204 (1981).
\newblock \doi{10.1016/0550-3213(81)90465-X}

\bibitem{berline1982classes}
Berline, N., Vergne, M.: Classes caract{\'e}ristiques {\'e}quivariantes.
  formule de localisation en cohomologie {\'e}quivariante.
\newblock CR Acad. Sci. Paris \textbf{295}(2), 539--541 (1982)

\bibitem{Bobev:2015kza}
Bobev, N., Bullimore, M., Kim, H.C.: {Supersymmetric Casimir Energy and the
  Anomaly Polynomial}.
\newblock JHEP \textbf{09}, 142 (2015).
\newblock \doi{10.1007/JHEP09(2015)142}

\bibitem{Bobev:2020egg}
Bobev, N., Charles, A.M., Hristov, K., Reys, V.: {The Unreasonable
  Effectiveness of Higher-Derivative Supergravity in AdS$_4$ Holography}.
\newblock Phys. Rev. Lett. \textbf{125}(13), 131,601 (2020).
\newblock \doi{10.1103/PhysRevLett.125.131601}

\bibitem{Bobev:2021oku}
Bobev, N., Charles, A.M., Hristov, K., Reys, V.: {Higher-derivative
  supergravity, AdS$_{4}$ holography, and black holes}.
\newblock JHEP \textbf{08}, 173 (2021).
\newblock \doi{10.1007/JHEP08(2021)173}

\bibitem{Bobev:2020pjk}
Bobev, N., Charles, A.M., Min, V.S.: {Euclidean black saddles and AdS$_{4}$
  black holes}.
\newblock JHEP \textbf{10}, 073 (2020).
\newblock \doi{10.1007/JHEP10(2020)073}

\bibitem{Bobev:2023dwx}
Bobev, N., David, M., Hong, J., Reys, V., Zhang, X.: {A compendium of
  logarithmic corrections in AdS/CFT}.
\newblock JHEP \textbf{04}, 020 (2024).
\newblock \doi{10.1007/JHEP04(2024)020}

\bibitem{Bobev:2022eus}
Bobev, N., Hong, J., Reys, V.: {Large N partition functions of the ABJM
  theory}.
\newblock JHEP \textbf{02}, 020 (2023).
\newblock \doi{10.1007/JHEP02(2023)020}

\bibitem{Boido:2022iye}
Boido, A., Gauntlett, J.P., Martelli, D., Sparks, J.: {Entropy Functions For
  Accelerating Black Holes}.
\newblock Phys. Rev. Lett. \textbf{130}(9), 091,603 (2023).
\newblock \doi{10.1103/PhysRevLett.130.091603}

\bibitem{Boido:2022mbe}
Boido, A., Gauntlett, J.P., Martelli, D., Sparks, J.: {Gravitational Blocks,
  Spindles and GK Geometry}.
\newblock Commun. Math. Phys. \textbf{403}(2), 917--1003 (2023).
\newblock \doi{10.1007/s00220-023-04812-8}

\bibitem{Butter:2010jm}
Butter, D., Kuzenko, S.M.: {New higher-derivative couplings in 4D N = 2
  supergravity}.
\newblock JHEP \textbf{03}, 047 (2011).
\newblock \doi{10.1007/JHEP03(2011)047}

\bibitem{Butter:2013lta}
Butter, D., de~Wit, B., Kuzenko, S.M., Lodato, I.: {New higher-derivative
  invariants in N=2 supergravity and the Gauss-Bonnet term}.
\newblock JHEP \textbf{12}, 062 (2013).
\newblock \doi{10.1007/JHEP12(2013)062}

\bibitem{Cassani:2019mms}
Cassani, D., Papini, L.: {The BPS limit of rotating AdS black hole
  thermodynamics}.
\newblock JHEP \textbf{09}, 079 (2019).
\newblock \doi{10.1007/JHEP09(2019)079}

\bibitem{Hristov-Cassia}
Cassia, L., Hristov, K.: {to appear}

\bibitem{Chester:2021gdw}
Chester, S.M., Kalloor, R.R., Sharon, A.: {Squashing, Mass, and Holography for
  3d Sphere Free Energy}.
\newblock JHEP \textbf{04}, 244 (2021).
\newblock \doi{10.1007/JHEP04(2021)244}

\bibitem{Couzens:2024vbn}
Couzens, C., L\"uscher, A.: {A geometric dual of F-maximization in massive type
  IIA}  (2024)

\bibitem{Dabholkar:2010uh}
Dabholkar, A., Gomes, J., Murthy, S.: {Quantum black holes, localization and
  the topological string}.
\newblock JHEP \textbf{06}, 019 (2011).
\newblock \doi{10.1007/JHEP06(2011)019}

\bibitem{Duistermaat:1982vw}
Duistermaat, J.J., Heckman, G.J.: {On the Variation in the cohomology of the
  symplectic form of the reduced phase space}.
\newblock Invent. Math. \textbf{69}, 259--268 (1982).
\newblock \doi{10.1007/BF01399506}

\bibitem{Ferrero:2020twa}
Ferrero, P., Gauntlett, J.P., Ipi\~na, J.M.P., Martelli, D., Sparks, J.:
  {Accelerating Black Holes and Spinning Spindles}  (2020)

\bibitem{Ferrero:2021etw}
Ferrero, P., Gauntlett, J.P., Sparks, J.: {Supersymmetric spindles}.
\newblock JHEP \textbf{01}, 102 (2022).
\newblock \doi{10.1007/JHEP01(2022)102}

\bibitem{Fuji:2011km}
Fuji, H., Hirano, S., Moriyama, S.: {Summing Up All Genus Free Energy of ABJM
  Matrix Model}.
\newblock JHEP \textbf{08}, 001 (2011).
\newblock \doi{10.1007/JHEP08(2011)001}

\bibitem{Geukens:2024zmt}
Geukens, S., Hong, J.: {Subleading analysis for $S^3$ partition functions of
  $\mathcal{N}=2$ holographic SCFTs}  (2024)

\bibitem{Gibbons:1976ue}
Gibbons, G., Hawking, S.: {Action Integrals and Partition Functions in Quantum
  Gravity}.
\newblock Phys. Rev. D \textbf{15}, 2752--2756 (1977).
\newblock \doi{10.1103/PhysRevD.15.2752}

\bibitem{Gold:2023ymc}
Gold, G., Hutomo, J., Khandelwal, S., Ozkan, M., Pang, Y.,
  Tartaglino-Mazzucchelli, G.: {All Gauged Curvature-Squared Supergravities in
  Five Dimensions}.
\newblock Phys. Rev. Lett. \textbf{131}(25), 251,603 (2023).
\newblock \doi{10.1103/PhysRevLett.131.251603}

\bibitem{Gold:2023ykx}
Gold, G., Hutomo, J., Khandelwal, S., Tartaglino-Mazzucchelli, G.: {Components
  of curvature-squared invariants of minimal supergravity in five dimensions}
  (2023)

\bibitem{Hatsuda:2016uqa}
Hatsuda, Y.: {ABJM on ellipsoid and topological strings}.
\newblock JHEP \textbf{07}, 026 (2016).
\newblock \doi{10.1007/JHEP07(2016)026}

\bibitem{Hristov-Hosseini-Reys}
Hosseini, S.M., Hristov, K., Reys, V.: {work in progress}

\bibitem{Hosseini:2019iad}
Hosseini, S.M., Hristov, K., Zaffaroni, A.: {Gluing gravitational blocks for
  AdS black holes}.
\newblock JHEP \textbf{12}, 168 (2019).
\newblock \doi{10.1007/JHEP12(2019)168}

\bibitem{Hristov:2021qsw}
Hristov, K.: {4d $ \mathcal{N} $ = 2 supergravity observables from
  Nekrasov-like partition functions}.
\newblock JHEP \textbf{02}, 079 (2022).
\newblock \doi{10.1007/JHEP02(2022)079}

\bibitem{Hristov:2022lcw}
Hristov, K.: {ABJM at finite N via 4d supergravity}.
\newblock JHEP \textbf{10}, 190 (2022).
\newblock \doi{10.1007/JHEP10(2022)190}

\bibitem{Hristov:2022pmo}
Hristov, K.: {The dark (BPS) side of thermodynamics in Minkowski$_{4}$}.
\newblock JHEP \textbf{09}, 204 (2022).
\newblock \doi{10.1007/JHEP09(2022)204}

\bibitem{Hristov:2023sxg}
Hristov, K.: {Explicit black hole thermodynamics in natural variables}.
\newblock JHEP \textbf{08}, 003 (2023).
\newblock \doi{10.1007/JHEP08(2023)003}

\bibitem{Hristov:2022plc}
Hristov, K.: {Maximally symmetric nuts in 4d \ensuremath{\mathscr{N}} = 2
  higher derivative supergravity}.
\newblock JHEP \textbf{02}, 110 (2023).
\newblock \doi{10.1007/JHEP02(2023)110}

\bibitem{Hristov:2023cuo}
Hristov, K.: {Black hole thermodynamics in natural variables: quadrophenia}.
\newblock JHEP \textbf{02}, 105 (2024).
\newblock \doi{10.1007/JHEP02(2024)105}

\bibitem{Hristov:2016vbm}
Hristov, K., Katmadas, S., Lodato, I.: {Higher derivative corrections to BPS
  black hole attractors in 4d gauged supergravity}.
\newblock JHEP \textbf{05}, 173 (2016).
\newblock \doi{10.1007/JHEP05(2016)173}

\bibitem{Hristov:2019mqp}
Hristov, K., Katmadas, S., Toldo, C.: {Matter-coupled supersymmetric
  Kerr-Newman-AdS$_4$ black holes}.
\newblock Phys. Rev. D \textbf{100}(6), 066,016 (2019).
\newblock \doi{10.1103/PhysRevD.100.066016}

\bibitem{Hristov:2018spe}
Hristov, K., Katmadas, S., Toldo, C.: {Rotating attractors and BPS black holes
  in $AdS_4$}.
\newblock JHEP \textbf{01}, 199 (2019).
\newblock \doi{10.1007/JHEP01(2019)199}

\bibitem{Hristov:2018lod}
Hristov, K., Lodato, I., Reys, V.: {On the quantum entropy function in 4d
  gauged supergravity}.
\newblock JHEP \textbf{07}, 072 (2018).
\newblock \doi{10.1007/JHEP07(2018)072}

\bibitem{Hristov-Yi-Gab}
Hristov, K., Pang, Y., Tartaglino-Mazzucchelli, G., et~al: {to appear}

\bibitem{Hristov:2021zai}
Hristov, K., Reys, V.: {Factorization of log-corrections in AdS$_{4}$/CFT$_{3}$
  from supergravity localization}.
\newblock JHEP \textbf{12}, 031 (2021).
\newblock \doi{10.1007/JHEP12(2021)031}

\bibitem{Hristov:2023rel}
Hristov, K., Suh, M.: {Spindle black holes in AdS$_{4} \times$ SE$_{7}$}.
\newblock JHEP \textbf{10}, 141 (2023).
\newblock \doi{10.1007/JHEP10(2023)141}

\bibitem{Hristov:2024qiy}
Hristov, K., Suh, M.: {Spindle black holes and theories of class $\mathcal{F}$}
   (2024)

\bibitem{Kallosh:1993wx}
Kallosh, R., Ortin, T.: {Killing spinor identities}  (1993)

\bibitem{Klare:2012gn}
Klare, C., Tomasiello, A., Zaffaroni, A.: {Supersymmetry on Curved Spaces and
  Holography}.
\newblock JHEP \textbf{08}, 061 (2012).
\newblock \doi{10.1007/JHEP08(2012)061}

\bibitem{Klare:2013dka}
Klare, C., Zaffaroni, A.: {Extended Supersymmetry on Curved Spaces}.
\newblock JHEP \textbf{10}, 218 (2013).
\newblock \doi{10.1007/JHEP10(2013)218}

\bibitem{Kuzenko:2022ajd}
Kuzenko, S.M., Raptakis, E.S.N., Tartaglino-Mazzucchelli, G.: {Covariant
  superspace approaches to ${\cal N}=2$ supergravity}  (2022)

\bibitem{Lauria:2020rhc}
Lauria, E., Van~Proeyen, A.: {${\cal N}=2$ Supergravity in $D=4,5,6$
  Dimensions}, vol. 966.
\newblock Springer (2020).
\newblock \doi{10.1007/978-3-030-33757-5}

\bibitem{LopesCardoso:1998tkj}
Lopes~Cardoso, G., de~Wit, B., Mohaupt, T.: {Corrections to macroscopic
  supersymmetric black hole entropy}.
\newblock Phys. Lett. B \textbf{451}, 309--316 (1999).
\newblock \doi{10.1016/S0370-2693(99)00227-0}

\bibitem{Marino:2011eh}
Marino, M., Putrov, P.: {ABJM theory as a Fermi gas}.
\newblock J. Stat. Mech. \textbf{1203}, P03,001 (2012).
\newblock \doi{10.1088/1742-5468/2012/03/P03001}

\bibitem{Martelli:2011fu}
Martelli, D., Passias, A., Sparks, J.: {The gravity dual of supersymmetric
  gauge theories on a squashed three-sphere}.
\newblock Nucl. Phys. B \textbf{864}, 840--868 (2012).
\newblock \doi{10.1016/j.nuclphysb.2012.07.019}

\bibitem{Martelli:2023oqk}
Martelli, D., Zaffaroni, A.: {Equivariant localization and holography}.
\newblock Lett. Math. Phys. \textbf{114}(1), 15 (2024).
\newblock \doi{10.1007/s11005-023-01752-1}

\bibitem{Nekrasov:2003rj}
Nekrasov, N., Okounkov, A.: {Seiberg-Witten theory and random partitions}.
\newblock Prog. Math. \textbf{244}, 525--596 (2006).
\newblock \doi{10.1007/0-8176-4467-9_15}

\bibitem{Nekrasov:2003vi}
Nekrasov, N.A.: {Localizing gauge theories}.
\newblock In: {14th International Congress on Mathematical Physics} (2003)

\bibitem{Nekrasov:2002qd}
Nekrasov, N.A.: {Seiberg-Witten prepotential from instanton counting}.
\newblock Adv. Theor. Math. Phys. \textbf{7}(5), 831--864 (2003).
\newblock \doi{10.4310/ATMP.2003.v7.n5.a4}

\bibitem{Nosaka:2015iiw}
Nosaka, T.: {Instanton effects in ABJM theory with general R-charge
  assignments}.
\newblock JHEP \textbf{03}, 059 (2016).
\newblock \doi{10.1007/JHEP03(2016)059}

\bibitem{Ooguri:2004zv}
Ooguri, H., Strominger, A., Vafa, C.: {Black hole attractors and the
  topological string}.
\newblock Phys. Rev. D \textbf{70}, 106,007 (2004).
\newblock \doi{10.1103/PhysRevD.70.106007}

\bibitem{Ozkan:2024euj}
Ozkan, M., Pang, Y., Sezgin, E.: {Higher Derivative Supergravities in Diverse
  Dimensions}  (2024)

\bibitem{Pestun:2007rz}
Pestun, V.: {Localization of gauge theory on a four-sphere and supersymmetric
  Wilson loops}.
\newblock Commun. Math. Phys. \textbf{313}, 71--129 (2012).
\newblock \doi{10.1007/s00220-012-1485-0}

\bibitem{Pestun:2016qko}
Pestun, V.: {Review of localization in geometry}.
\newblock J. Phys. A \textbf{50}(44), 443,002 (2017).
\newblock \doi{10.1088/1751-8121/aa6161}

\bibitem{Skenderis:2002wp}
Skenderis, K.: {Lecture notes on holographic renormalization}.
\newblock Class. Quant. Grav. \textbf{19}, 5849--5876 (2002).
\newblock \doi{10.1088/0264-9381/19/22/306}

\bibitem{Suh:2024asy}
Suh, M.: {Equivariant localization for wrapped M5-branes and D4-branes}  (2024)

\bibitem{deWit:2010za}
de~Wit, B., Katmadas, S., van Zalk, M.: {New supersymmetric higher-derivative
  couplings: Full N=2 superspace does not count!}
\newblock JHEP \textbf{01}, 007 (2011).
\newblock \doi{10.1007/JHEP01(2011)007}

\bibitem{deWit:2017cle}
de~Wit, B., Reys, V.: {Euclidean supergravity}.
\newblock JHEP \textbf{12}, 011 (2017).
\newblock \doi{10.1007/JHEP12(2017)011}

\bibitem{deWit:2006gn}
de~Wit, B., Saueressig, F.: {Off-shell N=2 tensor supermultiplets}.
\newblock JHEP \textbf{09}, 062 (2006).
\newblock \doi{10.1088/1126-6708/2006/09/062}

\bibitem{York:1986it}
York Jr., J.W.: {Black hole thermodynamics and the Euclidean Einstein action}.
\newblock Phys. Rev. D \textbf{33}, 2092--2099 (1986).
\newblock \doi{10.1103/PhysRevD.33.2092}

\end{thebibliography}

\end{document}